\documentclass[12pt]{iopart}

\def\hs{\,-\,}
\def\case#1/#2{\textstyle\frac{#1}{#2} }
\newcommand{\3}{{}^{(3)}}
\usepackage{graphicx}

\begin{document}
\title{Bounce behaviour in Kantowski-Sachs and Bianchi Cosmologies}
\author{Deon Solomons$^1$, Peter K. S. Dunsby$^{2,3}$ 
and George F. R.  Ellis$^2$} 
\address{$1.$ Cape Peninsula University of Technology, 
Cape Town, South Africa} 
\address{$2.$ Department of Mathematics and Applied Mathematics,
University of Cape Town, Rondebosch 7701, Cape Town, South Africa}
\address{$3.$ South African Astronomical Observatory, Observatory 
7925 Cape Town, South Africa.}
\date{\today}
\begin{abstract}
Many cosmological scenarios envisage either a bounce of the universe 
at early times, or collapse of matter locally to form a black hole 
which re\hs expands into a new expanding universe region. 
Energy conditions preclude this happening for ordinary matter 
in general relativistic universes, but scalar or dilatonic 
fields can violate some of these conditions, and so could 
possibly provide bounce behaviour. In this paper we show that 
such bounces cannot occur in Kantowski\hs Sachs models without 
violating the {\it reality condition} $\dot{\phi}^2\geq 0$. 
This also holds true for other isotropic spatially homogenous 
Bianchi models, with the exception of closed Friedmann\hs Robertson\hs Walker 
and Bianchi IX models; bounce behaviour violates the {\em weak energy condition} 
$\rho\geq 0$ and $\rho+p\geq 0$. We turn to the Randall-Sundrum type 
braneworld scenario for a possible resolution of this problem.
\end{abstract}
\section{Introduction}                       %
Already in the 1930's, Tolman \cite{tolman} proposed
that closed ($k=+1$) Friedmann\hs Robertson\hs Walker (FRW) universe
models might re\hs expand after collapsing towards a high
density state in the future; it was already known then that
this was not possible for $k=0$ and $k=-1$ models. This idea
of a ``phoenix universe'' has remained popular (see Dicke and
Peebles \cite{dipe} for a discussion), and has re\hs appeared recently
in new forms, for example specifically in Smolin's idea of collapse
to a black hole state resulting in re\hs expansion into a new expanding
universe region \cite{smolin} and more recently Easson and
Brandenberger \cite{easson} have discussed the general features
all such scenarios have in common. Thus it is interesting to investigate
under what conditions such a bounce might take place within the context
of General Relativity and other theories of gravity (see for example 
\cite{carloni}).

Essentially because of the Raychaudhuri equation \cite{ray,ehl,ell}, 
such bounces are not possible in FRW models if the active gravitational 
mass is positive: that is, if $\rho + 3p > 0$. On the other hand 
quantum fields and indeed classical scalar fields can violate this condition 
\cite{HE}. Hence such fields can in principle allow bounce behaviour 
in FRW models. However there are still other conditions that must be 
satisfied by such fields, for example the {\it reality condition} (RC)
$\dot{\phi}^2 \geq 0$ which is equivalent to requiring that the 
inertial mass density $\rho + p \geq 0$ 
(if this is not true, very anomalous physical behaviour can occur). 
Thus it is interesting to see when bounces can occur without violating 
such conditions.

It is not only FRW models that are of interest. The geometry of 
the universe at a bounce might be very different than the completely 
isotropic and spatially homogeneous FRW spacetimes; indeed  
because some spatially homogeneous modes are unstable \cite{we}, 
some more general geometry might be expected. Thus it is 
interesting to examine also what happens in models such as Bianchi 
models which are anisotropic and spatially homogeneous, in order to 
start exploring the full phase\hs space of possibilities. 
Additionally, in the scenario of black hole collapse and subsequent 
re\hs expansion, we might expect the geometry at the bounce to be that 
of a Kantowski\hs Sachs $k=+1$ model \cite{KS,ell67}, 
because this model has the same symmetries as the spatially homogeneous 
interior region of the extended (vacuum) Kruskal solution that 
represents the late stage of evolution of an isotropic black hole when 
the matter can be neglected. We may indeed expect that at late stages 
of collapse such a vacuum approximation will be valid, because in 
many Bianchi universes anisotropy dynamically dominates over matter 
at early and late times \cite{we}.

For these reasons we are interested in the bounce behaviour not only of 
FRW models but also of Bianchi and Kantowski\hs Sachs universes. In 
this paper, we prove a number of relevant results. Specifically, 
bounce behaviour in  LRS Bianchi types I (BI) and  III (BIII)  
or Kantowski\hs Sachs (KS) models with scalar fields violate 
the reality condition $\dot{\phi}^2=\rho+p\geq 0$ (we assume 
that the energy flux and anisotropic pressure vanish: 
$\pi_{\mu\nu}=0=q_{\mu}$ which follows for fields observed relative 
to the normal congruence of curves). For scalar field models 
with arbitrary self\hs interaction we prove a {\em no\hs bounce} 
theorem to this effect, viz. {\em bounce behaviour in LRS BI, BIII 
or KS models violates the reality condition} \footnote{It has been 
known for some time that flat and open FRW scalar field models
do not admit bounce solutions \cite{madsen}.}.  

We then present additional results that restrict bounce behaviour in 
other spatially homogeneous models to closed FRW and Bianchi type IX. 
Open FRW and all (un-tilted) spatially homogeneous models all violate 
the weak energy condition (WEC) $\rho\geq0$ {\em and} $\rho+p\geq 0$ at the 
time of a bounce. The strong energy condition (SEC) is violated in all 
such models, including flat FRW models. The RC and 
the null energy condition are violated in all spatially homogeneous 
models with the exception of Bianchi type IX and closed FRW models.

Finally we turn our attention to the braneworld scenario of 
Randall and Sundrum \cite{randall}. For a Kantowski-Sachs, BI or BIII 
metric confined to a single three-brane embedded in a 
five-dimensional bulk, a non\hs local anisotropic pressure and 
negative non\hs local energy density (e.g. {\em dark radiation}) 
projected from the bulk onto the brane via the bulk Weyl tensor modifies 
the Einstein field equations sufficiently for bounce behaviour to be 
consistent with the RC. 

We should also mention efforts to solve the standard cosmological 
puzzles such as the flatness, homogeneity and monopole problems 
via ekpyrotic models \cite{ekpyrotic} in the braneworld context as 
potential alternatives to inflation despite being plagued with 
their own difficulties. In particular there has been considerable 
discussion in the literature on the evolution of cosmological perturbations 
through a bounce in the ekpyrotic scenario (see \cite{ekpyrotic-perts}
for a representative sample). This issue although extremely interesting, 
does not directly related to main the aims of this paper.
\section{The Key Equations}\label{key}       %
In this section we set up the equations needed to investigate 
whether or not the dynamics of LRS BI, BIII and KS models admit 
bounce behaviour in the presence of a classical minimally 
coupled scalar field $\phi$ with a self\hs interaction 
potential $V(\phi)$. The metric tensor for these models can be 
written in the following form:
\begin{equation}
ds^2=-dt^2+X(t)^2dr^2+Y(t)^2(d\theta^2+S^2(\theta)d\phi^2)\;,\label{key.1}
\end{equation}
where
\begin{eqnarray}
S(\theta)=\left\{
\begin{array}{c}
\sin{\theta}~~ \textstyle{for}~~ k=+1\;,\cr
\theta~~ \textstyle{for}~~ k=0\;,\cr
\sinh{\theta}~~ \textstyle{for}~~ k=-1\;,
\end{array}
\right.\label{key.2}
\end{eqnarray}
$X(t)$ and $Y(t)$ are the expansion scale factors and we have 
chosen standard units where $8\pi G=c=1$.

Relative to a normal congruence of curves with tangent vector $u^{\mu}$, 
the energy\hs momentum tensor $T_{\mu\nu}$ for a scalar field 
takes the form of a perfect fluid (See \cite{we} 
page 17 for details):
\begin{equation}
T_{\mu\nu}=\rho u_{\mu}u_{\nu}+ph_{\mu\nu}\,,\label{key.3}
\end{equation}
with
\begin{equation}
\rho=\case{1}/{2}\dot{\phi}^2+V(\phi)\label{key.4}
\end{equation}
and 
\begin{equation}
p=\case{1}/{2}\dot{\phi}^2-V(\phi)\;.\label{key.5}
\end{equation}

It is easy to show that for the above metric and choice of $u^{\mu}$, 
the Einstein field equations can be written in terms of propagation
equations for the usual expansion $\Theta$, shear  
$\sigma^2=\frac{1}{2}\sigma^{\mu\nu}\sigma_{\mu\nu}$ and 3\hs curvature  
$\3 R$ scalars:
\begin{eqnarray}
&&\dot{\Theta}+\case{1}/{3}\Theta^2+2\sigma^2+\dot{\phi}^2-V(\phi)=0\;,
\label{field1}\\
&&\dot{\sigma}+\Theta\sigma-\case{1}/{2\sqrt{3}}\3 R=0\;,
\label{field2}\\
&&\3\dot{R}+\case{2}/{3}\Theta\3 R-\case{2}/{\sqrt{3}}\3 R\sigma=0\;,
\label{field3}
\end{eqnarray}
and the Gauss\hs Codazzi constraint
\begin{equation}
\3 R=\dot{\phi}^2+2V(\phi)+2\sigma^2-\case{2}/{3}\Theta^2\;,
\label{field4}
\end{equation}
where
\begin{equation}
\Theta=\frac{\dot{X}}{X}+\frac{2\dot{Y}}{Y}\;,~~~~\sigma=\frac{1}
{\sqrt{3}}\left(\frac{\dot{X}}{X}-\frac{\dot{Y}}{Y}\right)\;,
~~\3 R=\frac{2k}{Y^2}\;,
\end{equation}
and $k$ is a constant taking the values $k=0,-1,+1$ for BI, BIII and 
KS models respectively.
The above set of equations together with the Klein\hs Gordon equation
\begin{equation}
\ddot{\phi}+\Theta\dot{\phi}+\frac{\partial V}{\partial\phi}=0
\end{equation}
give a complete description of the dynamics of these universe models and
solutions may be obtained once $V(\phi)$ has been specified. 
The expansion may also be expressed in terms of a volume scale 
factor $a(t)$, viz. $\Theta = {3\dot{a}\over a}$.

It is worth mentioning that there is an alternate route to 
finding solutions by noticing that equations (\ref{field1}) 
and (\ref{field4}) can be combined to give expressions for the 
potential and momentum density of the scalar field:
\begin{eqnarray}
V(\phi)&=&\case{1}/{3}(\dot{\Theta}+\Theta^2+\3 R)\;,\\
\dot{\phi}^2&=&-\case{2}/{3}\dot{\Theta}-2\sigma^2+\case{1}/{3}\3 R\;,
\label{reality}
\end{eqnarray}
so that in principle $V(\phi)$ can be determined by first specifying the 
scale factor dependence $X(t)$ and $Y(t)$ and then running the field 
equations backwards to determine the potential. This method has been
used to generate exact FRW inflationary solutions \cite{madsen}. 

The aim here is not to attempt to find exact solutions but examine 
whether it is possible to find solutions that exhibit bounce 
behaviour subject to the RC: $\dot{\phi}^2\geq 0$. 
Hence we need a precise definition of what a bounce is, together 
with a careful examination of equation (\ref{reality}). 
This is discussed in the next section.
\section{The No\hs bounce theorem in          %
 BI, BIII and KS models}\label{c}             %
\subsection{Definition of a Bounce} \label{ca}%
Defining the expansion parameters 
\begin{equation}
x=\dot{X}/{X}\;,~~~ y=\dot{Y}/{Y}\;,
\end{equation} 
a bounce in $X$ occurs at time 
$t=t_0$ iff $x(t_0)=0$ and $\dot{x}(t_0)>0$ while a bounce in 
$Y$ occurs at time $t=t_1$ iff $y(t_1)=0$ and $\dot{y}(t_1)>0$. 
It is clear that although it may be possible to have a
bounce in one of the scale factors but not the other, 
this does not lead to a new expanding universe region. 
We therefore require that a bounce occurs both in $X$ and $Y$ 
scale\hs factors, even though they may in general occur at different times. 
\subsection{The reality condition at a bounce}
Let us now determine whether the RC 
for the scalar field is satisfied at a bounce. In order
to do this we need to write equation (\ref{reality}) in
terms of $x$ and $y$ and their derivatives. This is 
most easily done by first substituting for the spatial 
curvature $\3 R$ using equation (\ref{field2}):
\begin{equation}
\dot{\phi}^2=-\case{2}/{3}\dot{\Theta}-2\sigma^2+\case{2}/{\sqrt{3}}
\dot{\sigma}+\case{2}/{\sqrt{3}}\Theta\sigma\;.
\end{equation}
Now using 
\begin{equation}
\sigma=\case{1}/{\sqrt{3}}(x-y)\;,~~~~\Theta=x+2y\;,\label{variables}
\end{equation}
we obtain
\begin{equation}
\dot{\phi}^2=-2(\dot{y}+y^2-xy)\;.
\label{reality2}
\end{equation}
We can now state and prove the main result of this section 
\footnote{It has been brought to our attention that 
Toporensky and Ustiansky \cite{toporensky} has given an almost 
identical proof in the appendix to their paper. However, much more clarity
has been achieved by using the covariant equations (6-9) to derive this result.}. 
\\\\
\noindent {\bf Theorem:} Bounce behaviour in LRS Bianchi type I, III 
and Kantowski\hs Sachs models dominated by a minimally coupled 
scalar field is not permitted unless the RC $\dot{\phi}^2\geq 0$ is violated. 
\\\\
\noindent {\bf Proof:} 
The proof follows immediately from equation (\ref{reality2}) 
when evaluated for a bounce in the $Y$ direction at
time $t=t_1$. In this case $y(t_1)=0$ and $\dot{y}(t_1)>0$ so 
(\ref{reality2}) simplifies to give $\dot{\phi}(t_1)^2=-2\dot{y}(t_1)<0$, 
so that even if a bounce occurs in the $X$ direction it is only
possible in the $Y$ direction if the RC is violated.

\section{No bounce behaviour in other         %
 Bianchi models}\label{d}                     %
We now examine the extent to which energy conditions preclude 
bounce behaviour in isotropic, spatially homogeneous Bianchi models 
of class A and B. We refer the reader to Wainwright and Ellis \cite{we} 
for an interesting review and classification of Bianchi type universes. 
Wald's no-hair theorem for global anisotropy can be found in \cite{wald}.

\noindent Not all Bianchi type models isotropize at large times. 
Specifically, those types which do not admit FRW solutions become 
highly anisotropic \cite{collins}. Models that do admit FRW solutions; 
Bianchi I and $VII_0$ admits k=0 solutions, Bianchi V and $VII_h$ 
admits $k=-1$ solutions, and Bianchi IX admits $k=+1$ solutions. 
Type $VII_h$ will in general {\em not} approach isotropy, while the 
type IX models will recollapse after a finite time.

From the outset it is clear that bounce behaviour in the volume 
scale factor $a(t)$ violates the SEC. This can be 
seen from the Raychaudhuri equation for spatially homogeneous cosmologies 
\begin{equation} 
\dot{\Theta}+\textstyle{{1\over 3}}\Theta^2+ 2\sigma^2  
=  -\textstyle{{1\over 2}}(\rho+3p)\;, \label{d.1}
\end{equation}
which simplifies to 
\begin{equation}
-\textstyle{{1\over 2}}(\rho+3p)=\dot{\Theta}(t_0)
+ 2\sigma^2(t_0)>0\;, \label{d.2}
\end{equation} 
since $\Theta=0$ at the bounce in the volume scale factor, and 
positivity results from $\sigma^2$ being non-negative, together with 
the requirement that $\dot{\Theta}>0$ at the bounce. 
Hence $\rho+3p<0$, violating the SEC at the bounce.

In addition to this, bounce behaviour in spatially homogeneous 
models with negative spatial curvature $\3 R$ violates the WEC 
{\em $\rho\geq 0$ and $\rho+p\geq 0$}. To show 
this we use the Friedmann constraint for spatially homogeneous 
cosmologies \footnote{See Wainwright and Ellis \cite{we}.}
\begin{equation}
\rho+\sigma^2=\textstyle{{1\over 3}}\Theta^2+\textstyle{{1\over 2}}\3 
R\;, \label{d.3}
\end{equation} 
which simplifies to
$\rho+\sigma^2=\textstyle{{1\over 2}}\3 R<0$ at the bounce, 
implying that $\rho<0$, since $\sigma^2$ is non-negative. This 
violates the WEC in class A types  II, $VI_0$, $VII_0$ (except for 
a very special case) and VIII. Bounce behaviour can be obtained 
in type IX models, which can have positive curvature. In class B, 
the spatial curvature is negative for all models, hence the 
WEC is violated in all class B models.

Finally, spatially homogeneous cosmologies satisfy the RC 
provided that 
\begin{equation}
\dot{\phi}^2=\textstyle{{1\over 3}}\3 R-\textstyle{{2\over 3}}
\dot{\Theta}-2\sigma^2\geq 0\label{d.4}\ .
\end{equation}
This constraint (c.f equation 13) is obtained by subtracting equations (\ref{d.1}) 
and (\ref{d.3}), and then rearranging terms. Bounce behaviour implies 
that $\textstyle{{2\over 3}}\dot{\Theta}+2\sigma^2>0$, since the 
shear $\sigma^2$ is non-negative. Hence we can write condition (\ref{d.4}) 
in a somewhat weaker from, viz. {\em Models satisfying the RC 
necessarily have $\3 R>0$.} Therefore, spatially homogeneous 
cosmologies with $\3 R\leq 0$ violate the RC. This does not mean 
models with positive spatial curvature necessarily satisfy the RC, since 
they have to meet the stronger condition (\ref{d.4}), in cases where 
the shear is non-vanishing at the time of the bounce. However, for 
isotropic models, it is necessary and sufficient that the spatial 
curvature is positive. 

Bianchi class A types II, $VI_0$ and VIII all have negative spatial 
curvature, and for type I it is zero. Type $VII_0$ is negative or zero. 
Hence all these cases violate the RC, by virtue of the above argument. 
Type IX violates the RC for cases in which the spatial curvature is 
negative or zero.

In class B, since types IV, V, $VI_h$ and $VII_h$ have negative 
spatial curvature. Bianchi III is the  special case of $h=-1$ in 
Bianchi $VI_h$, and the {\em exceptional} case 
Bianchi $VI^*_{-1/9}$ has $h=-1/9$. They all violate the RC.   

\section{Bounce behaviour on a                %
Kantowski-Sachs brane}\label{e}               %
Recent ideas in string and M -theory \cite{Witten,Shiro,Lukas} have 
led to the notion that gravity could be a higher dimensional theory, 
that becomes 4-dimensional at low energies. In the Randall-Sundrum 
scenario \cite{randall} gravity can be localised on a 3-brane while 
a fifth dimension may remain non-compact. Observers are bound to a 
brane, which may have a more general metric than the induced Minkowski 
metric that Randall and Sundrum \cite{randall} assumed. A covariant 
geometric approach was first given by Shiromizu, Maeda and Sasaki 
\cite{Shiro} and also by Maartens \cite{Maartens}. We shall 
assume that a scalar field $\phi$ is bound to the brane, and that 
and higher-dimensional modifications of the standard Einstein 
field equations are imprinted via (i) local quadratic energy-momentum 
corrections that arise from the extrinsic curvature, 
and (ii) non\hs local effects from the free gravitational field 
in the bulk, transmitted via a projection of the bulk Weyl tensor 
onto the brane \cite{Maartens}.   

In five dimensions, the field equations \cite{randall,Maartens} are Einstein's 
equations, with a (negative) bulk cosmological constant $\widetilde{\Lambda}$ 
and the brane energy\hs momentum providing the source:
\begin{equation}
\widetilde{G}_{AB} =
\widetilde{\kappa}^2\left[- \widetilde{\Lambda}\widetilde{g}_{AB}
+\delta(\chi)\left\{
-\lambda g_{AB}+T_{AB}\right\}\right]\,. \label{1.1}
\end{equation}
The tildes denote the bulk (5\hs dimensional) generalisation of
standard general relativity quantities, and
$\widetilde{\kappa}^2= 8\pi/\widetilde{M}_{\rm p}^3$, where
$\widetilde{M}_{\rm p}$ is the fundamental 5\hs dimensional Planck
mass, which is typically much less than the effective Planck mass
on the brane, $M_{\rm p}=1.2\times 10^{19}$ GeV. The brane is
given by $\chi=0$, so that a natural choice of coordinates is
$x^A=(x^\mu,\chi)$, where $x^\mu=(t,x^i)$ are spacetime
coordinates on the brane. The brane tension is $\lambda$, and
$g_{AB}=\widetilde{g}_{AB}-n_An_B$ is the induced metric on the
brane, with $n_A$ the spacelike unit normal to the brane. Matter
fields confined to the brane make up the brane energy\hs momentum
tensor $T_{AB}$ (with $T_{AB}n^B=0$).

The modification to the standard Einstein equations \cite{Maartens}, 
with the new terms carrying bulk effects onto the brane are
\footnote{Note that we set $\kappa^2=8\pi G=c=1$}
\begin{equation}
G_{\mu\nu}=-\Lambda g_{\mu\nu}+T_{\mu\nu}+\widetilde{\kappa}^4S_{\mu\nu} 
- {\cal E}_{\mu\nu}\,.
\label{1.2}
\end{equation}
If $\lambda={6\over\widetilde{\kappa}^4}$, then the energy 
scales are related as follow:
\begin{equation}
\Lambda={1\over 2}\widetilde{\kappa}^2\left(\widetilde{\Lambda}
+{1\over 6}\widetilde{\kappa}^{2}\lambda^2\right)\; .
\end{equation}
The tensor $S_{\mu\nu}$ represents local quadratic energy\hs momentum 
corrections of the matter fields, and the ${\cal E}_{\mu\nu}$ is the 
projected bulk Weyl tensor transmitting non\hs local gravitational degrees 
of freedom from the bulk to the brane. All the bulk corrections may be 
consolidated into effective total energy density $\rho^{tot}$, 
pressure $p^{tot}$, anisotropic stress $\pi^{tot}_{\mu\nu}$ 
and energy flux $q^{tot}_{\mu}$. The modified Einstein 
equations take the standard Einstein form with a redefined energy\hs 
momentum tensor:
\begin{equation}
G_{\mu\nu}=-\Lambda g_{\mu\nu}+T^{\rm tot}_{\mu\nu}\,,
\label{25aa}
\end{equation}
where
\begin{equation}
T^{\rm tot}_{\mu\nu}= T_{\mu\nu}+\widetilde{\kappa}^{4}S_{\mu\nu}
-{\cal E}_{\mu\nu}\,.
\label{25bb}
\end{equation}
\subsection{Matter description on the Kantowski-Sachs brane}\label{ss:brane}
\label{ss:2.2}
A homogeneous scalar field $\phi(t)$ on the brane has an 
energy\hs momentum tensor given by equations (\ref{key.3})-(\ref{key.5}), 
and in the absence of local anisotropic pressure and energy 
flux $\pi_{\mu\nu}=0=q_{\mu}$, effective total energy\hs momentum tensor has
\begin{eqnarray}
\rho^{\rm tot} &=& \rho+{\rho^2\over 2\lambda}+\rho_*\,, \label{22c}\\
p^{\rm tot} &=& p+\rho(p+{\rho\over 2\lambda})+\case{1}/{3}\rho_* \label{22d} 
\end{eqnarray}
with a non\hs local energy density $\rho_*=-{\cal E}_{\mu\nu}u^{\mu}u^{\nu}$. 
A thorough analysis of the dynamics of FRW, Bianchi I and V models obeying 
a barotropic equation of state on the brane was done by Campos 
and Sopuerta \cite{Campos} and for a scalar field by 
Goheer and Dunsby \cite{GD}. A Kantowski-Sachs brane is 
characterised by vanishing acceleration, vorticity and non\hs local energy 
flux (similar to that for a Bianchi I brane, but with positive 3-curvature; 
see for instance \cite{Maartens,MSS}): 
\begin{equation} 
A^{\mu}=\omega^{\mu}=q^{\mu}_*=0\;, 
\end{equation}
hence the total energy flux vanishes
\begin{equation}
q_{\rm tot}^\mu =0 \,.
\label{22f}\end{equation}
However, the anisotropic stress does not necessarily 
vanish: \[\pi^{\rm tot}_{\mu\nu} =\pi^*_{\mu\nu}\,. \]
Here $\pi^*_{\mu\nu}=-(h_{\mu}^{\ \alpha}h_{\nu}^{\ \beta}-{1\over 3}
h^{\ \alpha\beta}h_{\mu\nu}){\cal E}_{\alpha\beta}$. 
There is no evolution equation for $\pi^*_{\mu\nu}$ 
on the brane, since the non\hs local anisotropic stress carry 
bulk degrees of freedom that cannot be determined by brane observers. 
In general, the projection of the five-dimensional field equations 
onto the brane, together with the $Z_2$ symmetry, does not lead 
to a closed system. The standard conservation equation of 
general relativity holds on the brane, together with a simplified 
propagation equation for the non\hs local energy 
density $\rho_*$ \cite{Maartens,MSS}: 
\begin{eqnarray}
&&\dot{\rho}+\Theta(\rho+p)=0\,,\label{22g}\\ &&
\dot{\rho}_*+{\textstyle{4\over3}}\Theta{\rho}_*
+\sigma_{\mu\nu}\pi_*^{\mu\nu}=0\,, \label{22h}
\end{eqnarray}
where $\Theta$ is the expansion parameter. Both $\rho_*$ and $\pi^*_{\mu\nu}$ 
are homogeneous on the brane. 
\subsection{Gravitational  Equations}\label{ss:2.3}
The Einstein Field equations (\ref{field1})-(\ref{field4}) 
of section \ref{key} are modified: 
\begin{eqnarray}
&&\dot{\Theta}+ {\textstyle{1\over 3}}\Theta^2+2\sigma^2
+{\textstyle{1\over 2}}(\rho+3p)
=\Lambda-{\textstyle{1\over 2}}(2\rho+3p)\case{\rho}/{\lambda}
-\rho_* \;,\label{23g}\\ 
&&\dot{\sigma}+\Theta\sigma-{\textstyle{1\over 2\sqrt{3}}}\3 R
=\case{1}/{2}\case{\sigma_{\mu\nu}\pi_*^{\mu\nu}}/{\sigma}\,,\label{23h}\\ 
&&\3 \dot{R}+{\textstyle{2\over 3}}\Theta\3 R
-{\textstyle{2\over\sqrt{3}}}\3 R\sigma= 0 \,,\label{23i}\\
&&{\textstyle{2\over 3}}\Theta^2+\3 R-2\sigma^2=2\Lambda+
2\rho\left(1+{\textstyle{\rho\over 2\lambda}}\right)+2\rho_*\,.\label{23j}
\end{eqnarray}
The effects of embedding the free gravitational field in the 
five\hs dimensional bulk are represented by the correction terms to the 
energy\hs momentum tensor $T_{\mu\nu}$, as expressed in 
equations (\ref{23g}), (\ref{23h}) and (\ref{23j}). 
\subsection{Feasible Bounce Behaviour}\label{ss:2.4}
If we subtract equation (\ref{23j}) from $2$ $\times$ 
equation (\ref{23g}):
\begin{equation} 
(1+\case{\rho}/{\lambda})(\rho+p)=\case{1}/{3}\3 R 
-\case{2}/{3}\dot{\Theta}-2\sigma^2-\case{4}/{3}\rho_*
\end{equation}
and then substitute $\3 R$ using equation (\ref{23h}), 
together with new variables $x$ and $y$ given 
by equations (\ref{variables}), we obtain
\begin{equation}
(1+\case{\rho}/{\lambda})(\rho+p)=-2(\dot{y}+y^2-xy)-
\case{4}/{3}\rho_*-\frac{1}{\sqrt{3}}{\sigma_{\mu\nu}
\pi_*^{\mu\nu}\over\sigma}\,.
\label{23k}
\end{equation}
The quantity 
$\case{1}/{\sqrt{3}}\sigma_{\mu\nu}\pi_*^{\mu\nu}/\sigma$ 
is generally non-zero. We then assume the induced metric for 
the brane is given by equations (\ref{key.1}) and (\ref{key.2}), 
which apply to Kantowski-Sachs, Bianchi I or III models. 
(The bulk metric for a Kantowski-Sachs or Bianchi braneworld is 
not known \cite{MSS}). It is evident that a bounce in $Y$ 
(see the definition in Section (\ref{ca})) becomes feasible by 
virtue of (i) the anisotropic pressure contracted with shear 
term, or (ii) in the absence of this contraction 
\footnote{In that case the non\hs local energy propagates 
as {\em dark radiation} $\rho_*=\rho_{*_0}(a/a_0)^{-4}<0$.}, 
provided the energy density is sufficiently negative. 

In order to examine what the conditions or a bounce to occur in $X$, 
we first substitute for $\3 R$ in equation (34) and then using (16) to 
replace the expansion and shear in terms of $x$ and $y$, we obtain  
\begin{eqnarray}
\dot{x} = -(x+2y)x+\case{1}/{2}(\rho-p-
\case{\rho p}/{\lambda})+\Lambda+\case{1}/{3}\rho_*
+\case{1}/{\sqrt{3}}{\sigma_{\mu\nu}\pi_*^{\mu\nu}\over\sigma}\;,
\end{eqnarray}
which when evaluated at a bounce leads to the condition
\begin{eqnarray}
\case{1}/{2}(\rho-p-\case{\rho p}/{\lambda})
+\Lambda+\case{1}/{3}\rho_*+\case{1}/{\sqrt{3}}{\sigma_{\mu\nu}
\pi_*^{\mu\nu}\over\sigma} > 0\;.\label{23l}
\end{eqnarray}
\subsection{Illustration}\label{ss:2.5}
In the case of a pure cosmological constant $\Lambda>0$
with $\sigma_{\mu\nu}\pi_*^{\mu\nu}=0$\footnote{In situations where $\pi^*_{\mu\nu}\sigma^{\mu\nu}\neq 0$, 
the conditions for a bounce in both $X$ and $Y$ may not be simultaneously satisfied, however a bounce
in the volume expansion ($\dot{\Theta}>0$) is still possible provided $\rho_*<0$.}, the reality condition 
is marginally satisfied, since, in this case equation (\ref{23k}) becomes
\begin{equation}
\dot{y}+y^2-xy=-\case{2}/{3}\rho_*\,.
\label{25a}
\end{equation}
If a bounce in scale factor $Y$ occurs at $y=0,\ \dot{y}>0$, 
then equation (\ref{25a}) implies that $\rho_*<0$.

Equation (\ref{23l}) gives
\begin{equation}
\dot{x}+x^2+2xy =\Lambda+\case{1}/{3}\rho_*\label{25c}
\end{equation}
and since a bounce in $X$ occurs at $x=0,\ \dot{x}>0$, we have 
\begin{equation}
-\Lambda<\case{1}/{3}\rho_*<0\,.\label{25d}
\end{equation}
This corresponds with the result of Santos {\em et al} \cite{MSS}, 
where it is shown that Bianchi branes are {\em stable} in this 
regime, but unstable if $\rho_*<-3\Lambda$. 
However, a bouncing Kantowski Sachs brane requires a stronger 
constraint. This is seen by casting equation (\ref{23j}) as 
\begin{equation}
\textstyle{{1\over 2}}\3 R = -(2x+y)y+\Lambda+\rho_*\,.\label{25e}
\end{equation}
Hence a bounce in $Y$ also means that $\case{1}/{2}\3 R=\Lambda
+\rho_*$, and since the 3-curvature is positive, 
$\rho_*>-\Lambda$. The KS brane bounce therefore constrains 
the non\hs local energy density to satisfy
\begin{equation}
-\Lambda<\rho_*<0\,.\label{25f}
\end{equation} 
\subsection{Phase Portrait}\label{ss:2.6}
In the Friedmann equation (\ref{23j}), we define the non-negative quantity
\begin{equation}
D^2={\textstyle{1\over 9}}\Theta^2+\textstyle{{1\over 6}}\3 R
-\case{1}/{3}\rho_* ={\textstyle{1\over 3}}\sigma^2 
+ {1\over 3}\Lambda\label{26a}
\end{equation}
and use this to define the dynamical variables:
\begin{equation}
Q^2={\Theta^2\over 9D^2},\ \ 
\Sigma^2={\sigma^2\over 3D^2},\ \ \tau=D t\, .\label{26b}
\end{equation} 
Clearly $Q,\Sigma\in[-1,1]$. There are three non-negative density parameters
\begin{equation}
\Omega_{\rho_*}:=-{\rho_*\over3D^2}, \ \ 
\Omega_{\Lambda}=:{\Lambda\over3D^2},\ \ \ 
\Omega_k={\3 R\over 6D^2}\label{26c}
\end{equation}
with the property that $\Omega_{\rho_*},\Omega_{\Lambda},\Omega_k\in[0,1]$. 
The Friedmann constraint (\ref{26a}) separates into two equations
\begin{equation}
\Sigma^2+\Omega_{\Lambda}=1,\ \ Q^2+\Omega_u+\Omega_k=1\;, \label{26d}
\end{equation}
hence the state space is compact. If we use the deceleration parameter
\begin{equation}q:=-{3\over \Theta^2}(\dot{\Theta}+{1\over 3}\Theta^2) 
\label{26e}\end{equation}
then the Raychaudhuri equation (\ref{23g}) becomes 
\begin{equation}
qQ^2=2\Sigma^2-\Omega_u-\Omega_{\Lambda}\, .
\label{26f}
\end{equation}
The non\hs local energy density equation (\ref{22h}), 
the shear and curvature evolution equations (\ref{23h}) and (\ref{23i}), 
together with (\ref{26d}) and (\ref{26e}) yield a closed system of ode's
\begin{eqnarray}
D' & = & (\Omega_k-3Q\Sigma)\Sigma D\label{26g}\\
Q' & = & -(1+q)Q^2-(\Omega_k-3Q\Sigma)\Sigma Q \label{26h}\\
\Sigma' & = & (1-\Sigma^2)(\Omega_k-3Q\Sigma) \label{26i}\\
\Omega_{\rho_*}' & = & -2\left[2Q+(\Omega_k-3Q\Sigma)\Sigma \right]
\Omega_{\rho_*} \label{26j}\\
\Omega_{\Lambda}' & = & -2(1-\Sigma^2)(\Omega_k-3Q\Sigma)\Sigma \label{26k}\\
\Omega_k' & = & -2\left[Q-\Sigma+(\Omega_k-3Q\Sigma)\Sigma\right]\Omega_k\;, 
\label{26l}
\end{eqnarray}
where is {\it prime} denotes differentiation w.r.t. $\tau$.

Since $D$ does not appear in equations (\ref{26h})-(\ref{26l}) 
we shall omit equation (\ref{26g}) from the rest of the 
discussion. The state space is now 5-dimensional. After 
replacing $q$, $\Omega_k$ and $\Omega_{\Lambda}$ from 
equations (\ref{26d}) and (\ref{26f}), the state space is 
reduced to three dimensions $(Q,\Sigma,\Omega_{\rho_*})$. 
\begin{eqnarray}
Q' & = & (1-Q^2)(1-Q\Sigma-3\Sigma^2)+(1+Q\Sigma)\Omega_{\rho_*} \label{26m}\\
\Sigma' & = & (1-\Sigma^2)(1-Q^2-3Q\Sigma-\Omega_{\rho_*}) \label{26n}\\
\Omega_{\rho_*}' & = & -2\left[2Q+(1-Q^2-3Q\Sigma-\Omega_{\rho_*})\Sigma)
\right]\Omega_{\rho_*}\,. 
\label{26o}
\end{eqnarray}
\begin{figure}
\begin{center}
\includegraphics{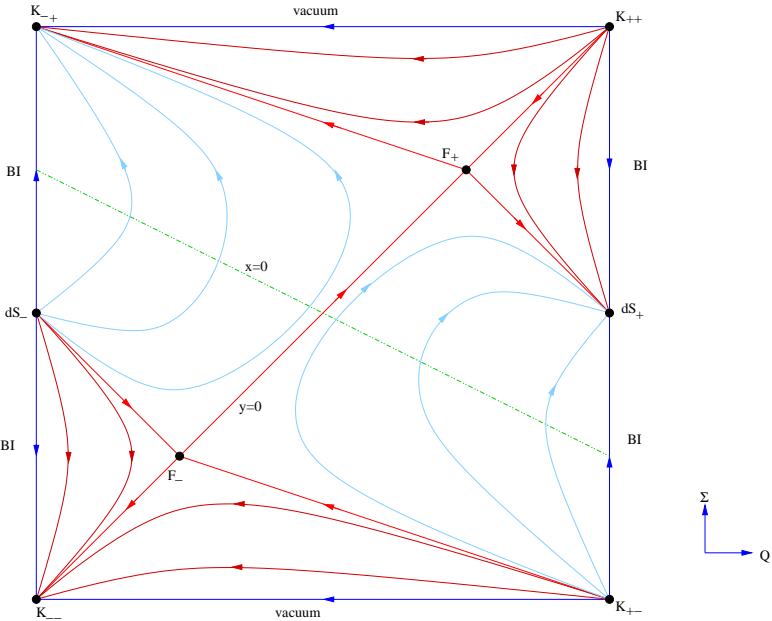}
\end{center}
\caption{\label{fig:figure1} State space {\cal R} is an invariant 
submanifold of {\cal M} that represents general relativity for 
$\3 R\geq 0$. The separatrix $y=Q-\Sigma=0$ precludes bounce behaviour 
in $Y$ from occurring.}
\end{figure}
\begin{figure}
\begin{center}
\includegraphics{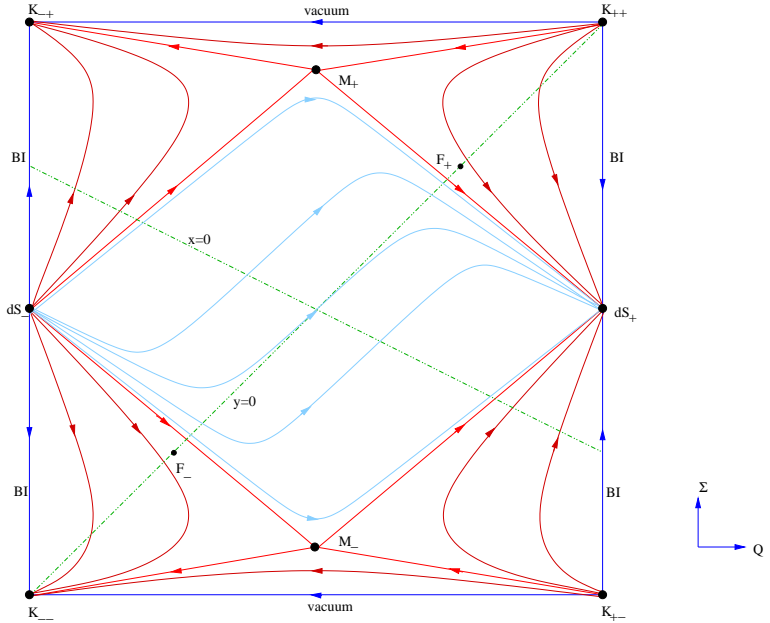}
\end{center}
\caption{\label{fig:figure2} State space {\cal M} no longer has 
a separatrix at $y=Q-\Sigma=0$ for $\Omega_{\rho_*}>0$. A {\em blue} 
trajectory that passes through the plane $x=Q+2\Sigma=0$ exhibits 
bounce behaviour in $X$, and a bounce in $Y$ when it passes through 
the plane $y=Q-\Sigma=0$.} 
\end{figure}
The boundaries $Q=\pm 1$ represent BI models with a 
cosmological constant, while the boundaries $\Sigma=\pm 1$ 
correspond to (vacuum) Kasner models in the invariant submanifold 
$${\cal R}=\{(Q,\Sigma,\Omega_{\rho_*}):|Q|\leq 1,|\Sigma|\leq 1,
\Omega_{\rho_*}=0\}$$ 
that represents general relativity. Table 1 contains a list of all 
critical points, their characterisation and eigenvalues. The points 
$K_{\pm\pm}$ and $K_{\pm\mp}$ are Kasner solutions that form the vertices 
of ${\cal R}$. The critical points $K_{++}$ and $K_{+-}$ are {\em repellers}, 
while $K_{--}$ and $K_{-+}$ are {\em attractors}. de Sitter spacetime $dS_{+}$
is an {\em attractor} and $dS_{-}$ (anti-de Sitter) is a {\em repeller} 
located on the respective boundaries $Q=1$ and $Q=-1$ of the submanifold 
${\cal R}$. There are two saddle points $F_{\pm}$ located on the separatrix 
$y=Q-\Sigma=0$ inside ${\cal R}$. The saddle points $F_{\pm}$ have 
$\{\3 R =2\Lambda,\ \Theta=\sqrt{3}\sigma=\pm\sqrt{\Lambda},\ \rho_*=0\}$. 
From Figure 1 it is clear that no trajectory emerging from $dS_{-}$ 
inside ${\cal R}$ can to reach $dS_{+}$ due to the separatrix $y=Q-\Sigma=0$. 
In Figure 2 we demonstrate how the presence of non-local density parameter 
$\Omega_{\rho_*}$ alters the picture. The saddle points $M_{\pm}$ 
represent static braneworld models with $\{\3 R=0,\ \
Theta=0,\ \sigma=\pm\sqrt{2\Lambda},\ 
\rho_*=-3\Lambda\}$. Since $\Omega_{\rho_*}=1$ these models 
are located inside the greater state space 
$${\cal M}=\{(Q,\Sigma,\Omega_{\rho_*}):|Q|\leq 1,|\Sigma|\leq 1,0
\leq\Omega_{\rho_*}\leq 1\}$$ 
but outside ${\cal R}$.  A trajectory emerging from $dS_{-}$ may 
now {\em exit} ${\cal R}$ and cross the plane $y=Q-\Sigma=0$ (which is 
no longer a separatrix for $\Omega_{\rho_*}>0$) as it evolves toward, and 
eventually away from the static models $M_{\pm}$, and enters $dS_{+}$. 
Such a trajectory passes  through $Q=0$ and close to $M_{\pm}$, indicative 
of bounce behaviour in the volume scale factor $a$, since $Q'$ remains 
positive.  
\begin{center}
\begin{tabular}{|c|c|c|} \hline
Model  & Coordinates   & Eigenvalues \\ \hline
$dS_{\pm}$ & $(\pm 1,\ 0,\ 0)$ & $(\mp 2,\ \mp 3,\ \mp 4)$\\
$K_{\pm\pm}$ & $(\pm 1,\ \pm 1,\ 0)$ & $(\pm 2,\ \pm 6,\ \pm 6)$ \\
$K_{\pm\mp}$ & $(\pm 1,\ \mp 1,\ 0)$ & $(\pm 2,\ \pm 2,\ \pm 6)$ \\
$F_{\pm}$  & $(\pm{1\over 2},\ \pm{1\over 2},\ 0)$ & $(\pm{3\over 2},
\ \mp 2,\ \mp 3)$ \\
$M_{\pm}$ & $(0,\ \pm\sqrt{{2\over 3}},\ 1)$ &  
$(\pm 2\sqrt{{2\over 3}},\ 2,\ -2)$ \\
\hline
\end{tabular}
\end{center}
Starting at $dS_{-}$  a trajectory passes through the 
plane $y=Q-\Sigma=0$, with $y$ changing sign from negative 
to $0$ to positive, and through the plane $x=Q+2\Sigma=0$ 
with $x$ changing sign from negative to $0$ to positive. It 
terminates at the attractor $dS_{+}$. Note that the bounce 
definition (\ref{ca}) requires the trajectory to cut both 
the $x$- and $y$-planes (the order may be reversed, depending 
on the path chosen), entering from the negative side of each 
plane, and exiting the positive side. If it passes through 
$(0,0,\Omega_{\rho_*})$ then the bounce  in $X$ and $Y$ is synchronous, 
i.e. the shear $\sigma$ and the expansion $\Theta$ both vanish there. 

\section{Conclusion}\label{f}                  %
We have shown that in General Relativity, LRS Bianchi type I, III 
and Kantowski\hs Sachs scalar field models that exhibit bounce 
behaviour violate 
the {\it reality condition} for the momentum density $\dot{\phi}^2\geq 0$. 

Thus Smolin's idea of collapse to a black hole state resulting 
in re\hs expansion into a new expanding universe region is not 
viable if the end\hs state of the universe after collapse into a black hole 
is described by a Kantowski\hs Sachs model, even in situations where the 
matter is dominated by a scalar field. 

None of the LRS Bianchi models besides closed FRW and Bianchi IX can 
support bounce behaviour without significant energy violations. 
We also analysed this phenomenon in the Randall-Sundrum type 
braneworld scenario. In this case there are non\hs local effects transmitted 
via the bulk Weyl tensor from the gravitational field in the bulk, 
onto the brane, that makes Kantowski-Sachs brane exhibit bounce 
behaviour, assuming that the required non\hs local effects
do indeed lead to a physical higher-dimensional bulk (an issue that 
still remains unresolved \cite{campus2}). Specifically, a Kantowksi-Sachs brane bounce 
requires negative non\hs local energy density 
if $\sigma_{\mu\nu}\pi_*^{\mu\nu}=0$, 
as demonstrated in section \ref{e}.   
\section*{References}                          %

\end{document}